\documentclass[aps,prb,showpacs,reprint]{revtex4-1}
    \usepackage{float}
    \usepackage{natbib}
    \usepackage{amsmath}
    \usepackage{graphicx,xcolor,xhfill,calc,xspace,}
    \usepackage{amssymb}
    \usepackage{tikz,pgfplots}
    \usepgfplotslibrary{groupplots}
    \usepackage{standalone}
    \usetikzlibrary{plotmarks}
    \usetikzlibrary{arrows}
    \usepackage[mathscr]{euscript}
    \newcommand{\ket}[1]{|#1 \rangle}

\begin{document}
    \title{Guided magnon transport in spin chains: transport speed and correcting for disorder  }
    \author{ Muhammad H.~Ahmed, Andrew D.~Greentree}
    \address{Chemical and Quantum Physics, School of Applied Sciences, RMIT University, Melbourne 3001, Australia}
    \date{\today}
    
    \begin{abstract}
    High fidelity quantum information transport is necessary for most practical models of quantum computation. By analogy with optical wave guides, a spatio-temporally varying magnetic potential on a one dimensional spin chain can achieve high fidelity transport of spin excitations. By comparing different potential shapes, we establish the effects of potential shape on the fidelity and transport speed. We incorporate disorder into our model and show methods to minimise its effect on transport. Finally, we discuss implementations of our scheme in several accessible systems based on hydrogenic approximations.   
    \end{abstract}

    \maketitle
    
    \section{Introduction}

   Methods for high-fidelity quantum information transport are interesting for a number of reasons.  These span the fundamental questions of how quantum information spreads in complicated environments such as random walks \cite{Konno:2002tt,Aharonov:1993jj}, to more practical issues such as quantum photosynthesis \cite{Romero:2014jm}, and on-chip quantum communication in solid-state quantum computers \cite{Bose:2007ji}.    
    Here we concentrate on one aspect of quantum information transport (QIT), namely transport of information through the Heisenberg chain.  This problem has a rich history  \cite {Bloch:1930ch,ValinRodriguez:2003we,Osborne:2004dt,Itoh:2005dj,Balachandran:2008eu, Schirmer:2009dba,Topp:2009vx, Makin:2012vi,Petrosyan:2010ed} and is undergoing renewed investigation due to its importance for certain models of solid-state quantum computation, especially dopant in silicon approaches to quantum computation \cite{Kane:1998wh,OBrien:2001jr,Skinner:2002en,deSousa:2004du,Itoh:2005dj}. 
    
     The Heisenberg chain is a line of spin-$1/2$ particles or qubits.  In most approaches to Heisenberg chain information transport, the information is either explicitly or implicitly encoded into quantised spin excitations, termed magnons.  Within this context, the goal of magnonic QIT is to transmit magnons through a system with the highest fidelity in the shortest time.  It is also interesting to understand the potential to extrapolate classical magnonic devices \cite{Kruglyak:2010cy,Lenk:2011el,Chumak:2014iw} into the quantum regime, although we will not discuss that topic here.
    
\begin{figure}[tb] 
    \includegraphics[width=0.9\columnwidth,clip]{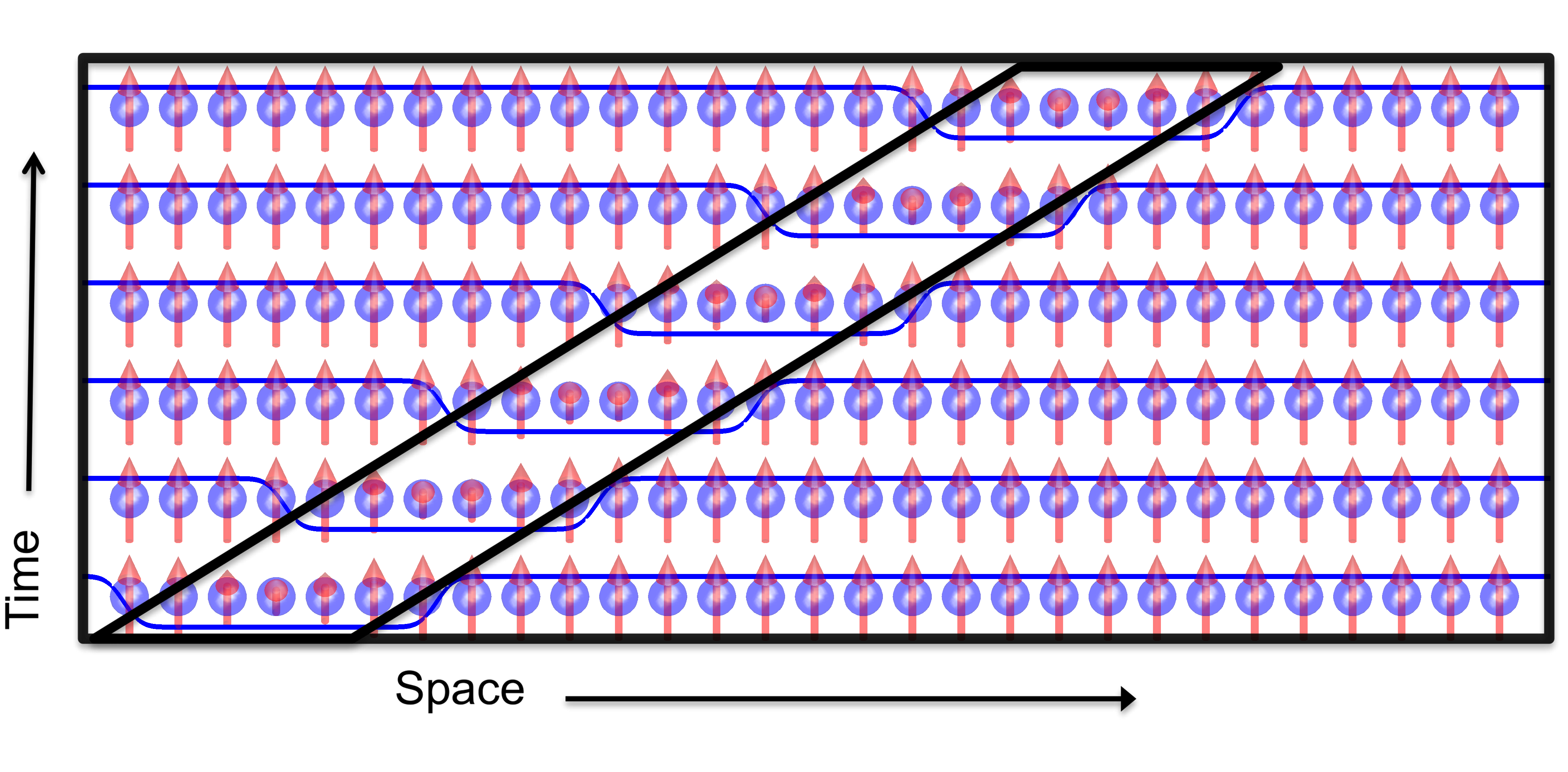}
    \caption{Snapshots showing the propagation of a magnon confined in a moving magnetic potential, indicated by the black lines, across a one-dimensional chain of spin $1/2$ particles. When the translation is adiabatic, the magnon remains in the local ground state.  We term the guiding potential a spin-guide.  A magnon propagating in a one-dimensional spin guide can be treated similarly to the propagation of a photon in a two-dimensional optical waveguide.}
    \label{fig1}
    \end{figure}

    When considering QIT in Heisenberg chains, there are several control approaches to consider.  The first is where there is no local control over the chain.  This regime has been considered by numerous authors and techniques for high-fidelity QIT typically involving precise timing \cite{Bose:2003dl,Heule:2011ki} or structuring of the chain \cite{Christandl:2003df}.  At the other extreme, one can consider complete local control, such as is envisaged in a completely controlled quantum computer, for example of the Kane style with nearest-neighbour coupling\cite{Kane:1998wh,Stephens:2007wt,Oh:2011ih}. There are also models for QIT based around control of just the ends of the chains, \cite{Greentree:2004di,Eckert:2007kl,Ohshima:2007ur,Schirmer:2009dba,Heule:2011ki}.     
    Between the limits of complete local control, or only end-of-chain control, there is at least one more regime, which we term \emph{semi-local} control.  In this regime, a confining potential is applied across the qubits that extends over a length that is large compared with the inter-qubit spacing, but small compared to the total spin-chain size, see Fig.~\ref{fig1}.  This regime has been considered previously \cite{Balachandran:2008eu,Makin:2012vi,Anonymous:L0FOM8O_} and under certain circumstances can be viewed as being the magnonic equivalent of a waveguide for light, which we term a \emph{spin-guide} \cite{Makin:2012vi}.  One of the main imperatives for studying such semi-local control is in the context of phosphorus in silicon quantum computing, where it is known that scalable control of qubits spaced at the 20~nm level is problematic given achievable control gate densities \cite{Copsey:2003ev,Rotta:2014vh}.

    Here we extend our previous analyses of spin-guides, explicitly showing the speed limits for QIT in spin-guides \cite{Makin:2012vi}, mechanisms to counteract disorder in the chain, and calculations for achievable spin-guide QIT in realistic media.  These calculations are performed for two sets of confining potentials, namely square-well and P\"{o}schl-Teller, although they can be generalised to any other potential.

    \section{Magnon propagation and the quantum speed limit}
    
    The Heisenberg spin chain for $N$ ferromagnetic spin 1/2 particles in a spatially and temporally varying magnetic field in the $z$ direction is
    \begin{align}
    H = -J \left[\sum_{n = 1}^{N} S_n^z S_{n+1}^z + \frac{1}{2}\left(S^+_n S^-_n+1 + h.c.\right)\right] - B_n(t)S^z_n, \label{eq:Ham}
    \end{align}
    
    \begin{figure}[tb] 
    \includegraphics[width=\columnwidth,clip]{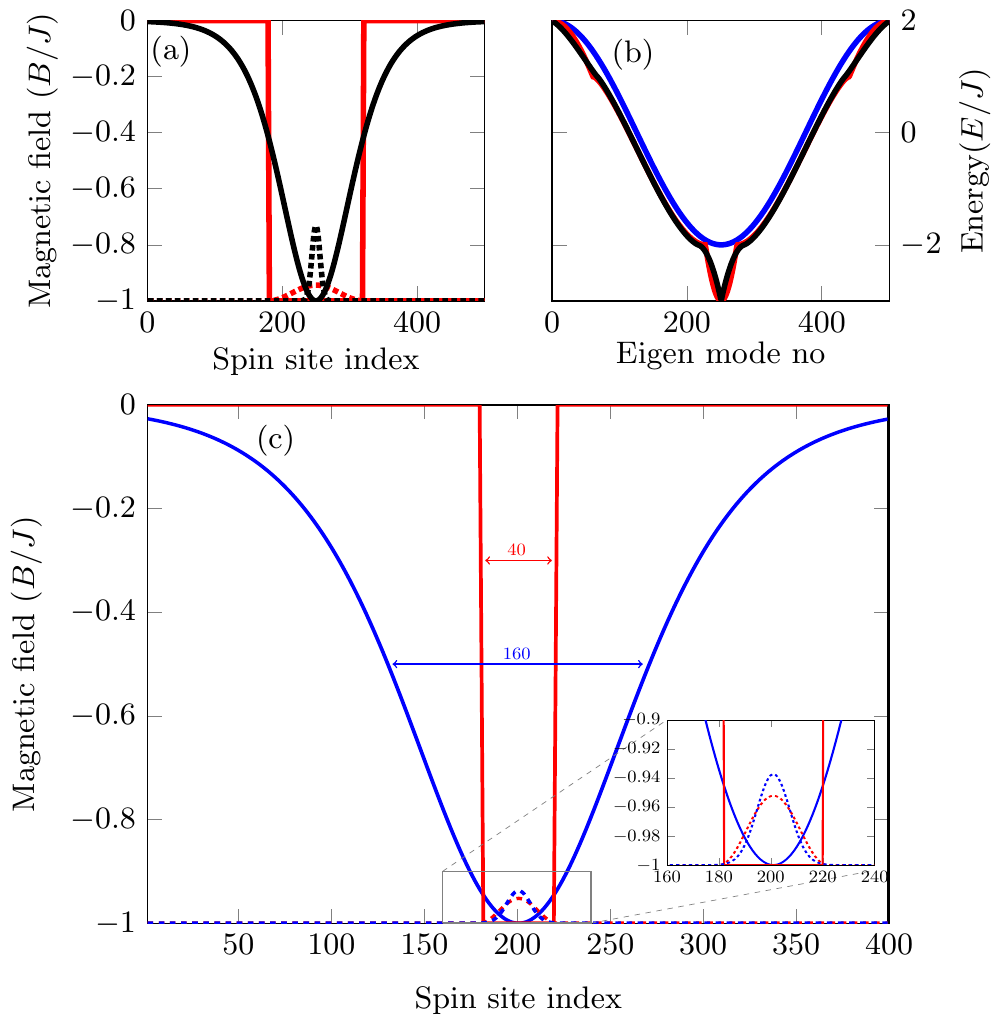} 
          \caption{Spin-guide confining potentials and their ground states. (a) Solid black and red lines showing the shape of equal width of P-T and SW potentials, respectively. The dotted lines show the corresponding ground states. (b) Eigenspectrum of a spin chain with no applied potential(blue). In the first excitation subspace, energy of Eigen modes varies between -2$J$ and 2$J$. Application of external field creates some modes with energy lower then -2$J$. These modes are bound modes. Black is the eigen energies of a spin chain with P-T magnetic field applied and red is with SW applied. (c) P-T (solid blue) and SW (solid red) potential, chosen such that the resulting ground state is of same width. Dotted blue is the P-T ground state and dotted red is SW ground state. For the same width of spin ground state, the P-T potential needs to be four times wider than the SW.}
  \label{pot_comp}
 \end{figure}

\noindent where $J$ is the exchange interaction strength, which is assumed to be isotropic ($ J_{x}=J_{y}=J_{z} =J$), $S_n^z$ is the $z$-Pauli matrix of the $n^{\text{th}}$ spin, and $S_n^+$ and $S_n^-$ are the spin raising and lowering operators for the $n^{\text{th}}$ spin.
    We maintain our system in the one-excitation subspace, and define the one-excitation basis states as

    \begin{align}
    \ket{n} \equiv \ket{\uparrow_n}\bigotimes_{m \neq n} \ket{\downarrow_m}, \quad \forall m,n,
    \end{align}
\noindent where the arrows denotes the spin projection with respect to the applied magnetic field and the subscript denotes the number of the spin along the chain. An arbitrary magnonic state in the one-excitation subspace can be written as

    \begin{align}
    \ket{\psi} = \sum_n c_n \ket{n},
    \end{align}

\noindent where $\sum_n |c_n|^2 = 1$.  More specifically, we are concerned with propagating magnonic states in one dimension, so we introduce the wavenumber $k$, and write down the propagating states as
    
    \begin{equation}
    \ket{\psi_k}=\sum_n e^{ikan}c_{n}\ket{n},
    \end{equation}
where $a$ is the lattice spacing and we have assumed that spin $n$ is at location $an$.    
Our treatment here is similar to that discussed in Ref.~\onlinecite{Makin:2012vi}, although here we have taken a discrete approach, rather than the continuum.  This discretisation allows us to address more realistic features of the propagation including disorder as the potential is swept along the spin chain, and to clarify the maximum speed limit for magnon propagation.

The process of spin guiding involves populating the ground state of the one-excitation subspace, and then adiabatically translating the potential, thereby moving the magnonic excitation. However, at least two minimum requirements can be stated.
    
    \begin{enumerate}
      \item The potential should have at least one non-degenerate bound mode.
      \item The momentum of the magnon should be well defined, and matched to the translation speed of the spin guide. A large magnon $k$-space spread prevents the matching of the entire wave function with the ground state of the spin guide. 
    \end{enumerate}
    To satisfy these requirement we chose the following:
    \begin{align}
    B_{\text{PT}}(x,t)&=B_0\text{sech}^{2}\left(\frac{x-x_0}{w}\right), \label{pt} \\ 
    B_{\text{SW}}(x,t)&=\frac{B_0}{2}\left[\text{erf}(x-x_0-w)+\text{erf}(x-x_0+w)\right], \label{sw}
    \end{align}
where $x_0 = x_0(t)$ is the center of the (moving) potential, and $w$ is a measure of the width of the confining potential in each case. $ B_{0} $ is the maximum depth of the potential well. In this way, a time-varying potential in one-dimension (i.e. a 1+1~D system) can be treated analogously to a two-dimensional waveguide type system\cite{Ladouceur:1996tk} (Fig \ref{fig1}). It is important to stress that this analogy allows most concepts of waveguide optics to be directly translated to the spin guide case. In particular  Ref.~\onlinecite{Makin:2012vi} showed the analogy of beamsplitter and interference, and any achievable refractive index profile should have a corresponding magnetic confining potential. The square well has an abruptly changing profile and therefore can be thought of as analogous to a step index waveguide. On the other hand, P-T is more akin to graded index fibre due to its slowly varying profile.

         \begin{figure}[tb] 
    \includegraphics[width=\columnwidth,clip]{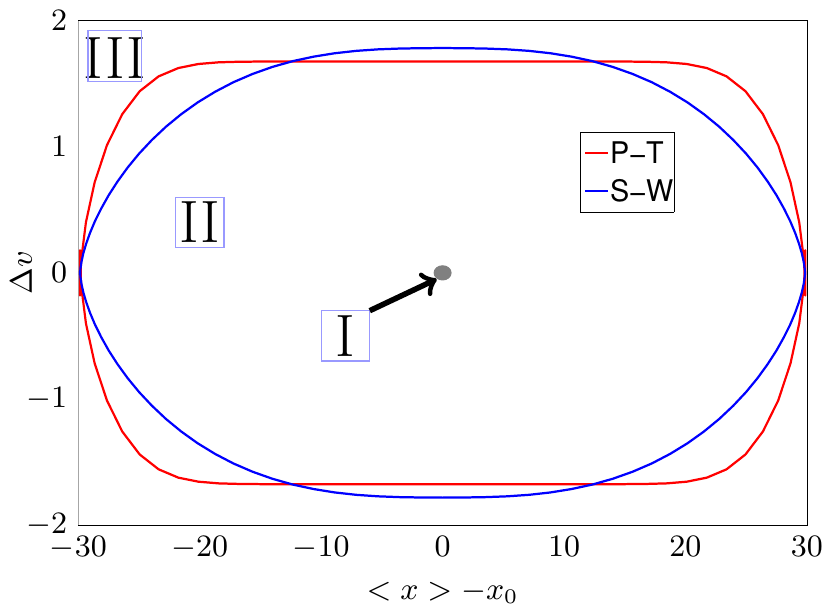} 
               	 \caption{Phase diagram showing several transport regimes, as a function of magnon speed and position with relative to the potential. $<x>$ is the expectation value of the magnon position and $x_0$ is the center of the confining potential, where $\Delta v$ is the speed mismatch between the magnon and the moving potential. This phase diagram was calculated by launching the magnon at increasing speed inside a static potential well ($w=60a$, $B_{0}/J=1$ ) and then calculating the instantaneous position and momentum of the magnon. If magnon and potential coincide with perfectly matched speeds then high fidelity quantum information transport will result (centre of the diagram, regime \textrm{I}). Small speed or position mismatch gives rise to coupling to higher modes and in general, loss of the phase information, but not population. In this case both position and momentum oscillate around the center (regime \textrm {II}). If the mismatch between position and/or speed is too great then the magnon will be coupled into unbound modes, leading to the loss of energy from the guided mode (regime  \textrm {III}).}
        	\label{sp_comp}
        	\end{figure}

    Three distinct regimes for guided magnon transport are shown in Fig.\ref{sp_comp}. The first regime is where the magnon is confined to the instantaneous ground state through out the propagation. This is the regime that is essential for the preservation of quantum phase, and hence for high-fidelity quantum information transport.  Regime \textrm{I} transfer can be achieved with any confining potential that matches the criteria above. The second regime is where the excitation is confined in the well, but excites to some nontrivial superposition of the other confined modes. This case is likely to be extremely sensitive to the precise details of transport. Hence we do not expect this regime to be useful for quantum information transport. Nevertheless, this regime may be useful where the transmitted information is the presence or absence of a magnon without phase information, for example the classical domain. Finally, there is the regime where the magnon is effectively unbounded and hence the spin guide is lossy.  As is shown in Fig \ref{sp_comp}, quantitatively, the SW and P-T confining potentials give similar results, with only a minor difference between the extent of the regimes. The boundary of regime \textrm{II} and regime \textrm{III} was calculated by launching a magnon in a static well with increasing speed until it starts to leaks out.

 \begin {figure*}  [th!] 
    \includegraphics[width=2\columnwidth,clip]{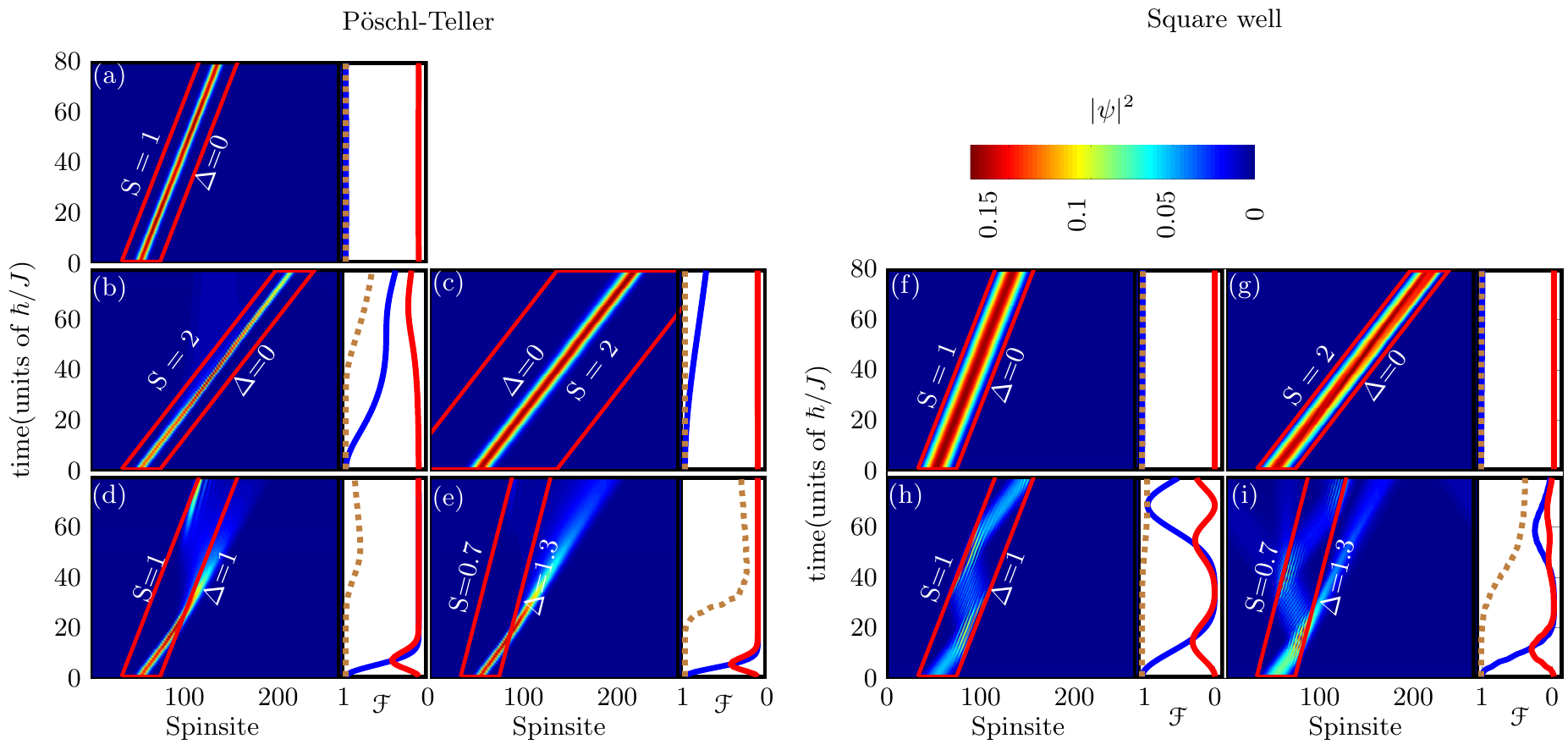} 
          \caption {Confined magnon transport using P-T and SW spin-guide [$B_{0}/J =1$ and $w =40a$ except for (c), where $w =160$]. The color axis show the probability density of confined magnon transport along the spin chain and the  solid red lines show the boundaries of the potential. S is the speed of the potential and $\Delta$ is the speed mismatch between magnon and potential. The system was initialised such that $v=S+\Delta$, where $v$ is the magnon group velocity. Graphs to the right of each plot show the ground state fidelity (solid blue), first excited state fidelity (solid red) and the confinement (dotted brown) of the magnon at each time instant. a, b, c, f and g are instances of regime \textrm{I} transfer where the initial magnon speed and potential speed were matched perfectly. However, (b) still show a partial loss in fidelity and confinement, which is due to a very localised ground state of the spin-guide. (c) is a repetition of (b) with an increased potential width, which shows a successful transfer. d and h show a regime \textrm{II} transfer, in which the initial speed of the magnon and the potential was differed by $\Delta$. In this case ground state fidelity goes to zero very quickly but the confinement still stays at one. e and i show the case when $\Delta$ becomes too large and both fidelity and confinement are lost.}
    \label{pt_fid}   
      \end{figure*}

    The magnon frequency and the group velocity in the one-dimensional Heisenberg spin chain is
    \begin{align}
    \label{gvel}
    \omega &= \frac{2J}{\hbar}\left[1-\cos(ka)\right], \\
    \label{vg}
       v_{g}&=\frac{\partial \omega}{\partial k}=\frac{2Ja}{\hbar}\sin(ka),
    \end{align}
    where $\omega$ is the frequency, and $ v_{g}$ is the group velocity of the wave packet. The appropriate $k$ for a given magnon speed can be determined to match the speed of the guiding potential, using eq \ref{vg}. It is evident from eq.~\ref{vg} that as k approaches $ \pi/2a $, $ v_{g}$ approaches its maximum of $2J$. So in a perfectly ordered system a magnon is bounded by a maximum speed  limit of $2J$ \cite{Balachandran:2008eu}.  \\

Figures \ref{pt_fid}a, \ref{pt_fid}b, \ref{pt_fid}c, \ref{pt_fid}f and \ref{pt_fid}g show instances where the speeds of the magnon and guiding potential are matched perfectly. Fidelity of the magnon transport was calculated by   
    \begin{equation}
    \mathscr{F}= |\langle e^{-ikx} \psi_{0}(t)|\phi(t)\rangle|^{2},
    \end{equation}
    
   \noindent where $|e^{-ikx}\psi_{0}\rangle $ is our ansatz for the instantaneous moving ground state of the spin-guide and $\phi(t)$ is the magnon wavefunction. Figure  \ref{pt_fid}b illustrates population loss due to a spread of momentum in the magnon.  The P-T potential has tightly localised the magnon, thereby increasing the spread of its momentum.  However, the confining potential can only be translated at a single velocity, and hence can only be matched to a single momentum component. Equivalent momentum matching for the P-T and SW potentials is achieved for a P-T width four times greater than that of a SW with the same depth. In Fig.~\ref{pt_fid}c, we show that high-fidelity transport can be achieved by making the potential wider.

Figures  \ref{pt_fid}d,  \ref{pt_fid}e,  \ref{pt_fid}h and  \ref{pt_fid}i show instances where the magnon-potential speed was not matched at the start of the protocol. The potential translates with the speed $S$ where the magnon was set to move with speed $v$, such that $v=S+\Delta$, where $\Delta$ is the magnon-potential speed mismatch. Figures  \ref{pt_fid}c and  \ref{pt_fid}d are examples of regime \textrm{II} like transfer, in which fidelity drops during the transport but the confinement does not drop. Where  Fig.~\ref{pt_fid}e and  \ref{pt_fid}i show instances when $\Delta$ becomes too large and the magnon couples to unbound modes, resulting in loss of confinement.

\begin{figure}[H] 
    \includegraphics[width=\columnwidth,clip]{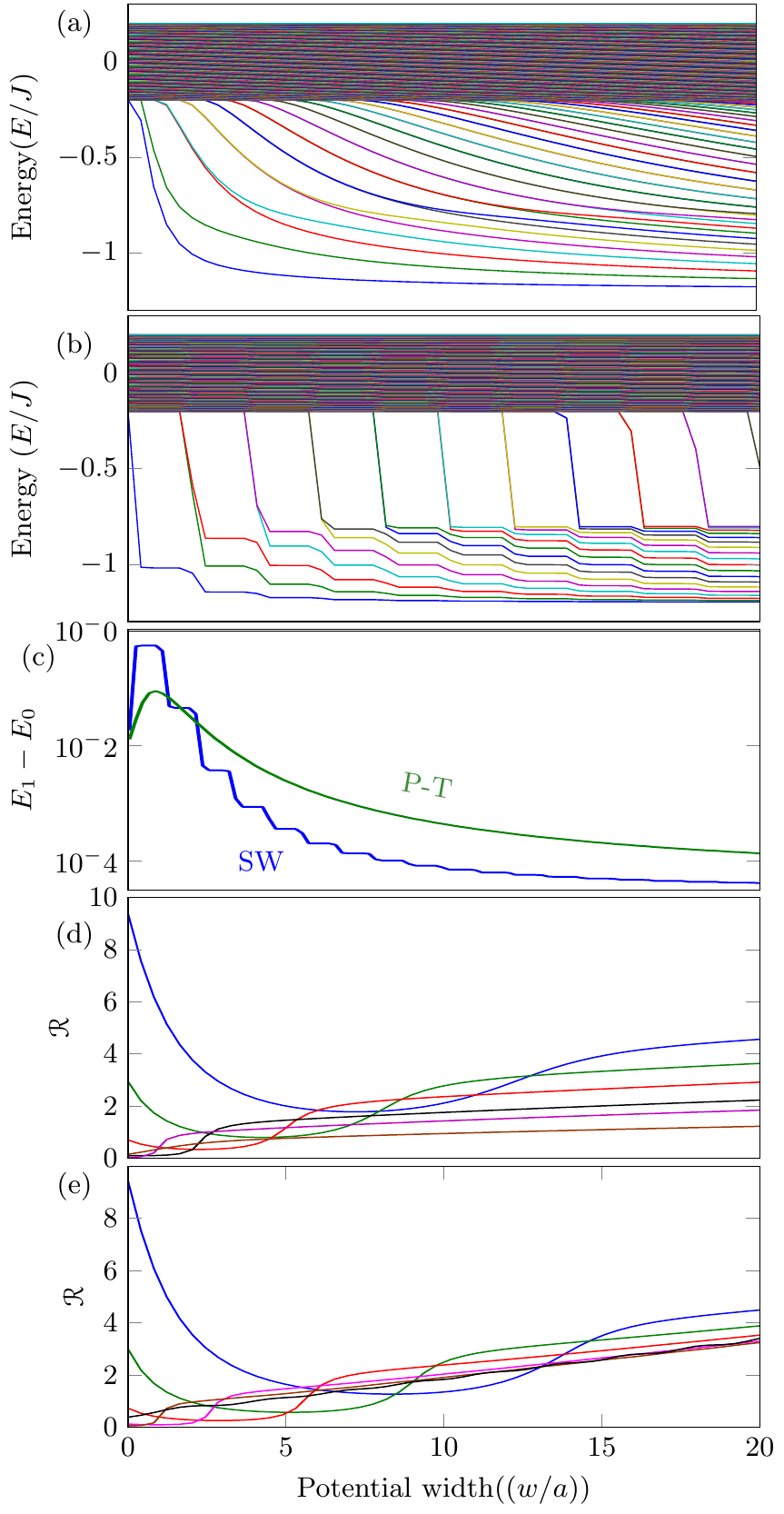}    
    \caption{Eigenspectra of (a) P-T and (b) SW as a function of the confining potential width for $B_{0}/J$=1. (c) First excitation energy gap of each potential.  As the width increases, unbound modes become bound modes. The energy gap between ground and first excited state is maximum when there is only one bound state. (e) and (f) show the reduced adiabaticity parameter ($\mathscr{R}$) of P-T and SW respectively. Each line represents different depths of the confining potential \textcolor{blue}{$B_{0}/J=0.05$(Blue)}, \textcolor[rgb]{0.13,0.55,0.12}{$B_{0}/J=0.1$(Green)}, \textcolor{red}{$B_{0}/J=0.2$(Red)}, $B_{0}/J=0.5$, \textcolor[rgb]{0.84,0.12,0.853}{$B_{0}/J=1$(Magenta)},  \textcolor{brown}{$B_{0}/J=5$(brown)}.}
    \label{width}
     \end{figure}

The confinement of a potential is directly proportional to the number of bound modes, and in turn the number of bound modes are dependent upon the shape of that potential. Figures.~\ref{width}a and \ref{width}b show the change in eigenspectrum of P-T and SW as a function of potential width, respectively, where Fig.~\ref{width}c shows the change in first excitation energy as function of potential width. As a limiting case of very small width, when the shape of potential approaches a delta function, there is always at least one bound state. As width increases, more and more unbound modes become bound modes by lowering their energy and also becoming spatially localised inside the potential. 
    
 Another useful quantity to understand the magnon propagation in spin-guides is the adiabiticity parameter, $A$. Adiabiticity is a measure of the probability that a magnon will make the transition from the ground to first excited state with a time varying Hamiltonian, with $A \ll 1$ indicating that the system prepared in an eigenstate will remain in that eigenstate.
    
    \begin{align}
    A&=\frac{\langle\psi_{0}|\partial_{t}H|\psi_{1}\rangle}{\arrowvert\langle\psi_{0}|H|\psi_{0}\rangle-\langle\psi_{1}|H|\psi_{1}\rangle|^{2}},
    \end{align} 
     
     \noindent where $ |\psi_{0}\rangle $ and and $ |\psi_{1}\rangle $  are the instantaneous ground and  first excited state respectively. Adiabiticity is an explicit function of speed, but implicitly it is also a function of the shape of the potential. The denominator terms are constant for a given potential shape and the numerator is a linear function of speed, which is manifested by $\partial_{t}H$ in the equation. As $A\propto$ spin guide speed, we define the reduced adiabaticity parameter, $\mathscr{R}$ as
\begin{align}
\mathscr{R}&\equiv\frac{A}{v}.  
\end{align}

To gain insight into the effects of the confining potential, we calculated  $\mathscr{R}$ for both potentials as the function of width and depth. It can be seen in Fig.~\ref{width}d and \ref{width}e  that $\mathscr{R}$ reaches its minimum when the first excitation gap is maximum. This is where there is only one bound mode in the guide and just before the second bound mode is formed. For both potentials, this condition occurs for $B_{0}/J=1$, approximately at width of $w=2 a$. At smaller potential well depths, the $\mathscr{R}$ minima occurs at larger widths. As an example, for $B_{0}=0.1J$ , $\mathscr{R}$ is minimised at $w=6 a$. After this minimum, $\mathscr{R}$ increases monotonically and linearly due to more unbound modes becoming bound. 
     
    \section{disorder}
   
\begin{figure}[t] 
    \includegraphics[width=\columnwidth,clip]{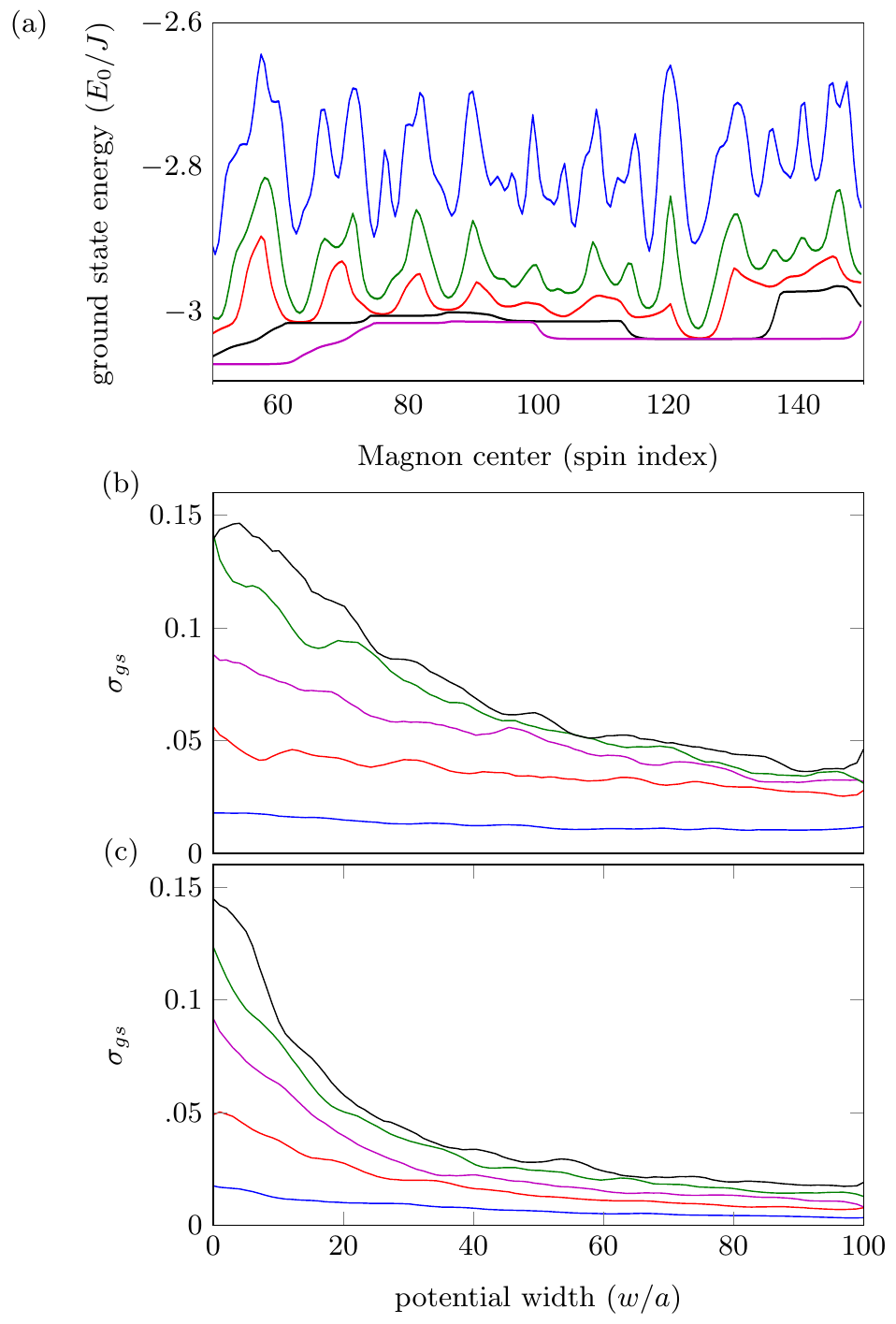}     
    \caption{(a) Magnon energy as a function of it's position along a disordered ($\sigma_{J}$ = 0.1) chain. We used a square well guide mean $B_{0}/J=1$. Each line represent different width of potential [\textcolor{blue}{$w=3(w/a)$(blue)}, \textcolor[rgb]{0.13,0.55,0.12}{$w=6(w/a)$(green)}, \textcolor{red}{$w=10(w/a)$(red)}, $w=25(w/a)$(black), \textcolor[rgb]{0.84,0.12,0.853}{$w=50(w/a)$(magenta)}]. Increasing the potential width has an averaging effect on these fluctuations and the magnon path becomes smoother, which is helpful in high fidelity  transport. (b) and (c) Standard deviation in the ground state energy ($\sigma_{gs}$) of a disordered spin chain, as potential moves across the chain for (b) SW and (c) P-T. Each line represent different $\sigma_{J}$ [\textcolor{black}{$\sigma_{J}=2\%$(Black)}, \textcolor[rgb]{0.13,0.55,0.12}{$\sigma_{J}=6\%$(Green)}, \textcolor[rgb]{0.84,0.12,0.853}{$\sigma_{J}=10\%$(Magenta)}, \textcolor{red}{$\sigma_{J}=14\%$(Red)}, \textcolor{blue}{$\sigma_{J}=18\%$(Blue)}]. Again, Increasing the spin guide width reduces the fluctuation in the ground state energy for both potentials.}
    \label{gs_disorder}
    \end{figure}

In realistic systems, spatial disorder of spin particles gives rise to the variations in the strength of the inter-spin coupling. This causes variations in the eigenspectrum as the potential sweeps across the spin chain, as shown in Fig~\ref{gs_disorder}a. These energy fluctuations give rise to the scattering, hence the increased possibility of magnon transition to excited states or trapping via Anderson localisation. Our goal is to smooth the fluctuations and thereby maintain the magnon in the moving ground state of the spin guide.

 \begin{figure*}[t] 
\includegraphics[width=2\columnwidth,clip]{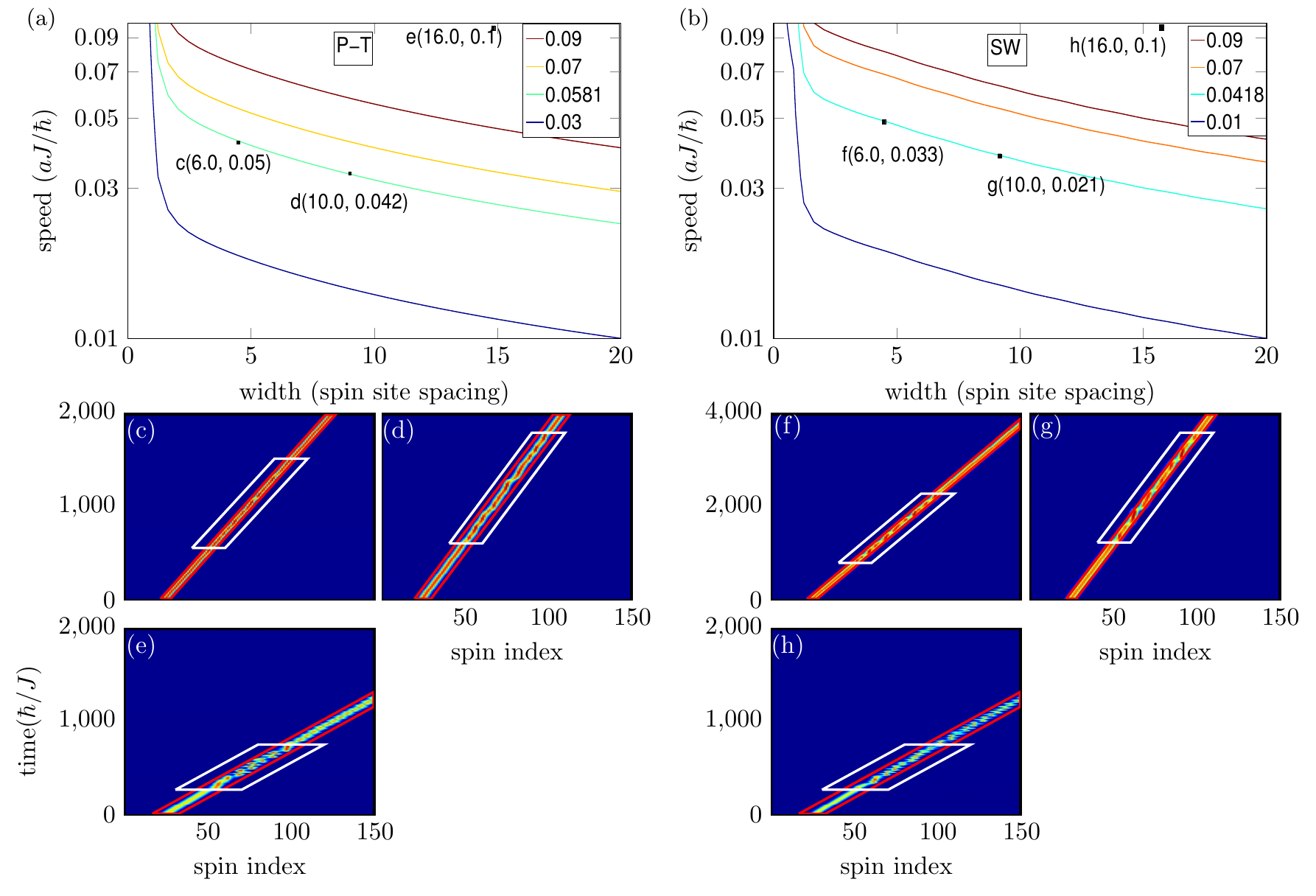} 
   \caption{Confined magnon transport on a spin chain that is perfect at the ends and has disorder in the middle. The magnon was initialised with the matching speed in the perfect part of the chain and then it was guided through the disordered part. If the magnon adiabiticity is below a certain threshold then it stays in the ground state. But, if the adiabiticity is above that threshold then it couples to higher modes and results in regime \textrm{II} like transfer. This adiabaticity threshold can be calculated empirically. (a) and (b) show the contours of constant adiabaticity as the function of width and speed for P-T and SW respectively, where the contour values are given in the legends. The color plots show the magnon transport using a guide with dimension corresponding to the points in (a) and (b), where white boxes show the disordered regions of the chain. In (c), (d), (f) and (g) the adiabaticity lies at the threshold line and the magnon stays in the ground state. (e) and (h) are the instances with adiabaticity higher then the threshold. Magnon get coupled to higher modes and results in regime \textrm{II} like transfer.}
  \label{dis_per}
  \end{figure*}

 Our approach is to increase the width of the spin guide. Fig.~\ref{gs_disorder}a shows the magnon energy in a disordered chain as a function of its position. As the potential moves along the chain, the energy varies, depending on the immediate environment of the potential well. However, increasing the potential width results in smoothing of the magnon energy and thereby minimising scattering. Fig.~\ref{gs_disorder}b and \ref{gs_disorder}c  show the standard deviation in the ground state energy, as the function of potential width. Each line represents different degrees of disorder. These results also show that as we increase the width of the potential, there is less fluctuation in the ground state energy. The disorder was implemented by randomly choosing the $J$ coupling with the probability of obtaining a particular value given by  (up to normalisation)
 \begin {equation}
P(J|J_{0},\sigma_{J}) =\begin{cases} e^{-\frac{(J - J_{0})^{2}}{2\sigma_{j}^{2}}} & \mbox{if }   \quad J_{0}-\sigma_j \leq J \geq J_{0}+\sigma_J\\
0 & \mbox{otherwise.} \end{cases}
 \end{equation}

 \noindent where $J_{0}$ is the mean $J$-coupling and $\sigma_{j}$ is the standard deviation of $J$-coupling. This probability distribution function was normalised by hand. For a chain with a finite disorder, we can calculate contours of constant adiabaticity in the space of width and speed, as shown in Fig.~\ \ref{dis_per}.

 For a chain with $\sigma_{J}=0.1$ , through repeated simulation we were able to empirically determine the minimum adiabaticity required for $\mathscr{F} > 0.99$, which is $A= 0.0581$ for P-T and $A=0.0418$ for square well with $B_{0}=1$ in each case. Fig.~\ref{dis_per} shows instances of confined transport through a chain that is perfect at the ends with disorder in the middle. We initialised the magnon in the moving ground state of the guiding potential in the ordered part of the chain, with matching speed and position. Then we guided it through the disordered part. When the magnon appears on the other side of the disordered part, depending on the adiabiticity, it can either still be in the ground state (regime \textrm{I}) or it can be coupled to the higher modes (regime \textrm{II}). Figures \ref{dis_per}c, d, f and g show instances where the adiabiticity is within the threshold limit and the magnon travels through the guide without being coupled to higher eigenstates. e and h are instances with adiabaticity higher than the threshold and the magnon is coupled to higher lying eigenstates, resulting in loss of phase information.

\section{Realistic Systems}

We now turn our attention to practical systems for achieving spin-guide transport.  We first discuss hydrogenic scaling laws, and then focus on specific material implementations. A hydrogenic approximation is ideal to quickly investigate the effects of changing spin site separation $a$ and potential width $w$. 

 \begin{figure}[t] 
    \includegraphics[width=\columnwidth,clip]{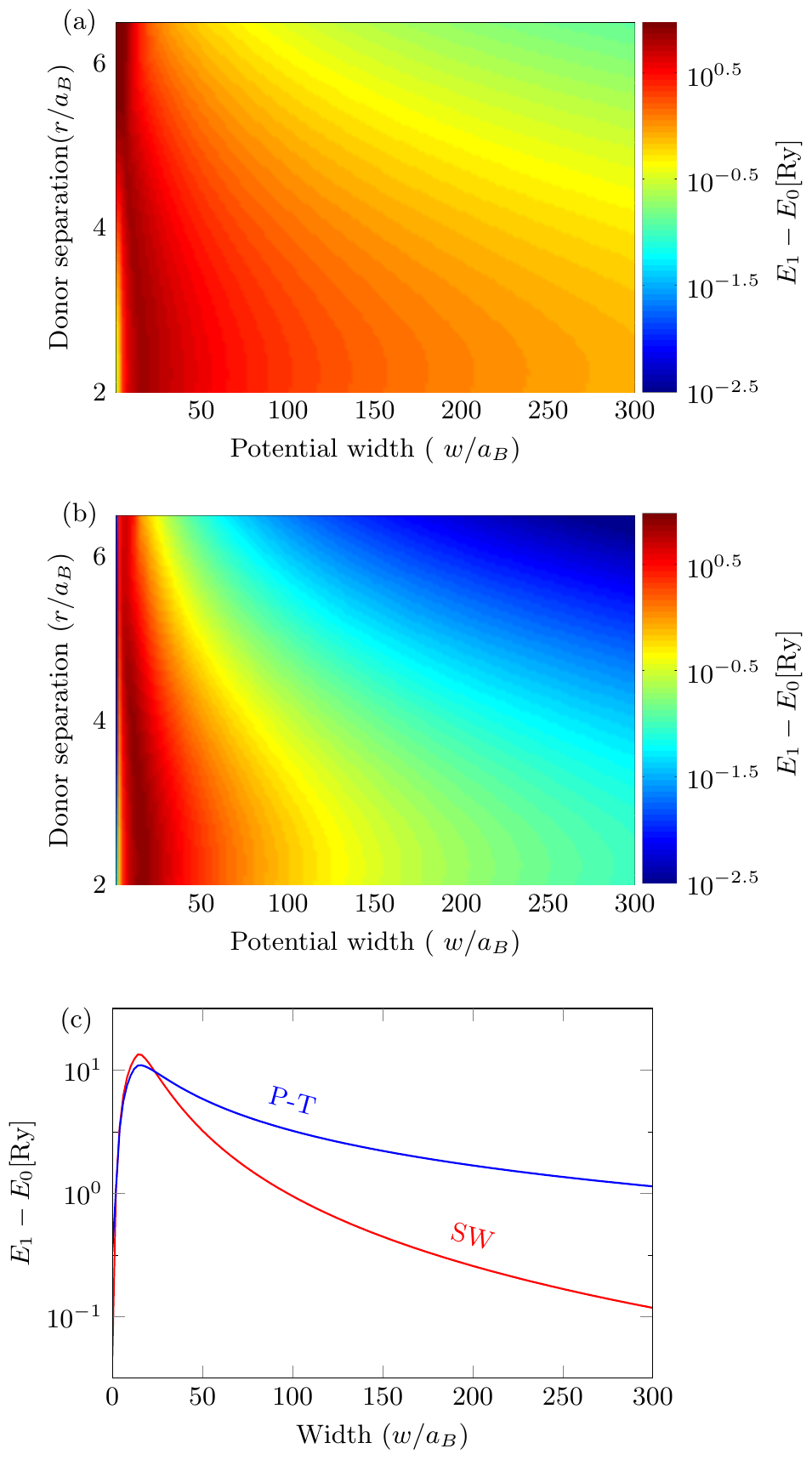} 
   \caption{Ground to first excited state energy gap for a hydrogenic Heisenberg spin chain, as a function of donor separation ($r$) and spin guide width ($w$) for (a) P-T spin guide and (b) SW spin guide, where both axis are normalised to $a_B$, the Bohr radius, and $B_0= 4.8\times10^{-3}$~Ry.  (c) Line slice of ground to first excited state energy separation when the inter donor spacing is $2a_B$ with $B_0/J=0.05$. }
    \label{ph_diag}
   \end{figure}
   
One metric for quickly evaluating the operation of a spin-guide is the ground to first excited state energy separation, with larger energy separations leading to increased robustness.  We calculated the first excitation energy gap as function of spin separation and potential width for the P-T and SW spin-guides, Figs.~\ref{ph_diag} a and b respectively. The potential width and donor separation were scaled in units of the Bohr radius and the energy gap scaled in units of the Rydberg constant appropriate for the system of interest. The $J$-coupling of donor atom as function of donor separation was calculated using \cite{Herring:1964cw} 
\begin{align}
J(r) = 0.4\frac{e^2}{a_B}\left(\frac{r}{a_B}\right)^{5/2}\exp\left(\frac{-2r}{a_B}\right),
\end{align}
where $a_B$ is the Bohr radius and $r$ is the separation between neighbour spins.  

Both phase plots suggest that the energy gap maximum shifts to wider well widths as the spin become closer. This dependence on inter-atomic spacing is to be expected as it gives rise to a strongly coupled chain.  

   \begin{table}[b]
    \centering
    \caption{Realistic one dimensional systems, their J-Coupling and maximum achievable speed. Si\textsuperscript{29} on Si\textsuperscript{28} is a chain of nuclear spins coupled through dipole-dipole coupling, where the other three systems are electronic spin chains coupled through exchange coupling.}
    \begin{tabular}{c c c c c c}
    \hline\hline
    System & J-Coupling  &             Spin        &     Max.  &        Max.    & Ref    \\ [0.5ex] 
                  &    [meV]      & Separation  &   Speed  &  speed   &    \\ [0.5ex] 
                  &		      & [\AA] & [m/s] & [sites/s] & \\
    \hline
    Co:Pt&20&20    & 0.6 &$3 \times 10^{8 }$ &  \onlinecite{Vindigni:2005cq}\\
    P:Si& 0.41& 93 &0.61 &$6.56 \times 10^{7} $& \onlinecite{Koiller:2002ih}\\
    P:Ge & 0.42 & 103 &0.65  &$6.31\times10^{7}$& \onlinecite{Koiller:2002ih}\\ 
    Si\textsuperscript{29}:Si\textsuperscript{28}&$1.0 \times 10^{-8}$&1.9 & $3.0\times 10^{-6} $&$1.57 \times 10^{4}$& \onlinecite{Ladd:2001fs}\\[1ex]
    \hline
    \end{tabular}
        \label{real}
    \end{table}

We also find that for donors separated by 2 $a_B$, the spin-guide can be relatively broad, up to 30~$a_B$ for SW and 15~$a_B$ for P-T if the energy gap is to maintained at 8~Ry.  This result supports our aim of achieving magnon guidance with semi-local control. To put such results in context, if we consider a phosphorus in silicon system, where phosphorus Bohr radius is $a_B \sim 3$~nm and lattice constant of silicon is 5.4~\AA. Then the Fig.~\ref{ph_diag}c corresponds to a system in which each donor is located at every 11th lattice site and the width of potential required to achieve $E_1-E_0=8$~Ry will be 90~nm for SW and 150~nm for the P-T. Generating confining potentials at such length scales in phosphorus in silicon is relatively straightforward. As with previous results, we find that the P-T potential has a larger energy gap than the SW for the same potential width and depth.  However both confining potentials are capable of magnon transport providing that the appropriate guiding speed is applied.

We now turn our attention to practical systems for the realisation of the spin guides, identifying four systems with good prospects for experimental demonstrations. These systems are cobalt on platinum \cite{Vindigni:2005cq}, phosphorus in silicon \cite{Koiller:2002ih}, phosphorus in germanium \cite{Koiller:2002ih}, and silicon 28 on silicon 29 nuclear spin chains \cite{Ladd:2001fs}.  The pertinent parameters for nearest neighbour couplings to realise an effective Heisenberg spin-chain are summarised in Table.~\ref{real}.

 Systems with high $J$-coupling to site separation ratio ($J/a$) give high maximum achievable speeds. The cobalt on platinum system has the highest achievable ratio of $J/a$ and so the highest maximum achievable speed. In Si\textsuperscript{29} in Si\textsuperscript{28}, the J-coupling is in fact nuclear dipole-dipole coupling. As this is many orders of magnitude smaller then exchange coupling \cite{Ladd:2001fs,Vindigni:2005cq}, the resulting magnon speed is commensurately lower.

    \section{conclusion}
    The central idea of this work is to model a scalable, solid-state, quantum communication protocol suitable for on-chip quantum communication, without the requirement for local qubit  control. We showed that a magnon can be adiabatically guided in a spin chain using a spatio-temporaly varying confining(magnetic) potential where the potential varies over length scales large compared with the inter-qubit space: semi-local control. We identified three different regimes of confined transport and compared the effect of different shapes and sizes of confining potential on their guiding properties. We found that a P\"oschl-Teller  is a better choice of quantum information transport than the more abruptly varying square well potential.  Such results are expected to apply when comparing any smoothly varying potential to any abruptly varying potential, and can be thought of as being analogous to the comparison between guidance properties for graded index vs step index optical waveguides.  As with optical waveguides, we also find that the magnon confinement for the square well potential is tighter than that of the P\"oschl-Teller.\\    
For a perfect system, 2$J$ is the maximum speed at which a magnon can travel in a one dimensional spin chain, guided or unguided. By considering the disorder in realistic systems we showed that a guided magnon transport is still achievable in disordered systems. The effects of disorder can be ameliorated by widening the spin-guide and effectively averaging over the disorder.  However, this comes at the cost of reduced energy separation between ground and first excited state, and hence slower magnon speeds are required for high-fidelity transport.\\
Our results have highlighted a technique for quantum information transport in one-dimensional spin chains.  Whilst we have focussed here on the practicalities of transport, it is important to recognise that there is a complete correlation between magnonic spin-guides and optical waveguides \cite{Makin:2012vi}.  Hence we expect that our techniques can be used to predict the operation of more complicated structures such as interferometers.  Whilst our results have considered single magnon propagation, they should also apply in the classical limit where many magnons might exist in the same spin-guide, and so have applicability to the growing field of magnonics.

 \section*{Acknowledgements}
The authors would like to thank Jackson Smith, Jan Jeske and Jared Cole for useful discussions.  This work was supported by the Australian Research Council (Grant No. DP130104381).


    \bibliography{myref}
    
    \end{document}